\documentclass[sigconf]{acmart}
  
\usepackage{bbm}
\usepackage{balance}
\usepackage{cleveref}
\usepackage{multirow}
\usepackage{graphicx}
\usepackage{enumitem}
\usepackage{xcolor}

\usepackage{algorithm,algpseudocode}%

\usepackage{bm}
\usepackage{color}
\usepackage{apxproof}
\usepackage{gensymb}
\usepackage{pifont}%

\usepackage{xspace}
\usepackage{mathtools}

\renewcommand\footnotetextcopyrightpermission[1]{}

\begin{document}

\title{TwinCL: A Twin Graph Contrastive Learning Model for Collaborative Filtering}

\author{Chengkai liu}
\affiliation{
  \institution{Texas A\&M University}
  \city{College Station, Texas}
  \country{USA}}
\email{liuchengkai@tamu.edu}

\author{Jianling Wang}
\affiliation{
  \institution{Texas A\&M University}
  \city{College Station, Texas}
  \country{USA}}
\email{jwang713@tamu.edu}

\author{James Caverlee}
\affiliation{
  \institution{Texas A\&M University}
  \city{College Station, Texas}
  \country{USA}}
\email{caverlee@tamu.edu}

\begin{abstract}

In the domain of recommendation and collaborative filtering, Graph Contrastive Learning (GCL) has become an influential approach. Nevertheless, the reasons for the effectiveness of contrastive learning are still not well understood. 
In this paper, we challenge the conventional use of random augmentations on graph structure or embedding space in GCL, which may disrupt the structural and semantic information inherent in Graph Neural Networks. Moreover, fixed-rate data augmentation proves to be less effective compared to augmentation with an adaptive rate. In the initial training phases, significant perturbations are more suitable, while as the training approaches convergence, milder perturbations yield better results. We introduce a twin encoder in place of random augmentations, demonstrating the redundancy of traditional augmentation techniques. The twin encoder updating mechanism ensures the generation of more diverse contrastive views in the early stages, transitioning to views with greater similarity as training progresses.
In addition, we investigate the learned representations from the perspective of alignment and uniformity on a hypersphere to optimize more efficiently. Our proposed Twin Graph Contrastive Learning model -- TwinCL -- aligns positive pairs of user and item embeddings and the representations from the twin encoder while maintaining the uniformity of the embeddings on the hypersphere. Our theoretical analysis and experimental results show that the proposed model optimizing alignment and uniformity with the twin encoder contributes to better recommendation accuracy and training efficiency performance. In comprehensive experiments on three public datasets, our proposed TwinCL achieves an average improvement of 5.6\% (NDCG@10) in recommendation accuracy with faster training speed, while effectively mitigating popularity bias. The code and datasets are available at \url{https://github.com/chengkai-liu/TwinCL}.

\end{abstract}

\ccsdesc[500]{Information systems~Recommender systems}

\keywords{Recommendation, Collaborative Filtering, Graph Neural Networks, Contrastive Learning}

\maketitle
\section{Introduction}

Recommender systems play a pivotal role in personalized information filtering, providing suggestions for items that are most pertinent to particular users~\cite{wang2021survey, wu2022graph, liu2024mamba4rec}. A fundamental personalized recommendation technique is collaborative filtering~\cite{sarwar2001item, he2017neural}, which models historical user-item interactions toward estimating future user preferences towards items. Recently, Graph Neural Networks (GNNs)~\cite{wang2019neural, he2020lightgcn, ying2018graph, sun2019multi} have gained prominence in collaborative filtering for their ability to capture indirect and non-obvious linkages between users and items, resulting in effective user and item representations in recommendation scenarios. Moreover, as contrastive learning has become widely used in recommender systems~\cite{yao2021self}, there has been a recent push to integrate contrastive learning with GNN-based recommenders~\cite{wujc2021self, yu2022graph}. These Graph Contrastive Learning (GCL) approaches for collaborative filtering are designed to tackle data sparsity to improve the performance of GNN recommenders.

While promising in certain cases, prior efforts to directly apply GCL models to recommendation systems still encounter significant challenges. Firstly, the prevalent use of augmentation techniques in GCL, such as edge or node dropping, can introduce unintended alterations to the underlying structural and semantic information of graphs. While effective in certain scenarios (e.g., citation graphs), these alterations could potentially undermine the performance of recommendation systems reliant on already sparse user-item interaction graphs (e.g., the sparsity of the Alibaba-iFashion dataset is 21 times that of the Cora dataset). For instance, inaccurately adding or dropping connections that do not reflect true similarity or dissimilarity relationships between users and items can distort their semantic meaning, resulting in inaccurate recommendations. Moreover, excessive node or dropout rates could obscure essential evidence regarding a user's preferences for specific items. %

Secondly, while GCL models have primarily focused on optimizing contrastive learning objectives, largely overlooking the potential benefits of \textit{alignment} and \textit{uniformity}~\cite{wang2020understanding, wang2022towards} in learned representations, previous GCL models for recommendation systems usually optimize the Bayesian personalized ranking (BPR)~\cite{rendle2012bpr} loss and the contrastive learning loss simultaneously. However, the necessity of negative sampling~\cite{yang2020understanding} in the BPR loss hinders the training efficiency due to the significant impact of negative samples' quality. Moreover, the BPR loss primarily focuses on pairwise relative ranking of items without explicit consideration for diversity, leading to popularity bias issues, especially in cases of data sparsity. Furthermore, contrastive learning with inappropriate perturbations may face challenges in capturing the signals of long-tail items, due to the bias introduced by the BPR loss.

To overcome these challenges, we propose in this paper a twin graph contrastive learning model for recommendation, called TwinCL. With respect to the first challenge, in pursuit of \textit{simplifying GCL for collaborative filtering}, we discard conventional random augmentations applied to the graph structure or embedding space. Inspired by momentum-updated mechanism~\cite{he2020momentum, jin2021multi} in CL, we adopt a strategy that performs contrastive learning using a momentum-updated twin encoder with the primary graph encoder. The twin encoder iteratively updates itself from the primary encoder, achieving a balance between historical parameter values and the current gradient of the primary encoder. This twin-based model has the benefit of preserving critical information in sparse user-item interaction graphs and efficiently refining representations with reduced computational operations. Additionally, it offers varied contrastive views during the initial stages, transitioning to more similar views in the later stages due to its updating mechanism, which is demonstrated to be effective in enhancing contrastive learning. With respect to the second challenge, we derive the alignment and uniformity losses from ~\cite{wang2022towards} and directly optimize the recommendation task \textit{without negative sampling}~\cite{yang2020understanding}. Through the optimization of alignment and uniformity properties within user and item embeddings, the training process attains faster convergence, while concurrently exhibiting reduced susceptibility to overfitting, thus enhancing robustness. Moreover, the design of alignment and uniformity, in conjunction with the twin encoder, synergistically contributes to achieving enhanced performance. In the initial few epochs, the model benefits from the optimization of alignment and uniformity, allowing the graph encoder to update to more suitable parameters. Through the acquisition of higher-quality parameters from the graph encoder, the twin encoder can iteratively refine its own parameters by drawing knowledge from the primary graph encoder, thereby generating superior node representations.

In summary, the main contributions of this paper are as follows:
\begin{itemize}
[leftmargin=*,noitemsep,topsep=1.5pt]
	\item We propose a new graph contrastive learning paradigm, TwinCL, which uses a momentum-updated twin encoder without any random augmentations on graph and node embeddings.
	\item We incorporate the optimization of alignment and uniformity with our twin graph contrastive learning model which improves the performance, efficiency, and robustness against popularity bias.
	\item We conduct extensive experiments on three public datasets, demonstrating that our proposed TwinCL consistently outperforms various competitive collaborative filtering baselines in terms of both recommendation accuracy and training efficiency.
\end{itemize}

\section{Preliminaries}

\subsection{Alignment and Uniformity}

In the domain of contrastive learning, \textit{alignment} and \textit{uniformity} have been identified as two key properties to measure the quality of representations~\cite{wang2020understanding}. Given a distribution of positive pairs $p_\text{pos}(\cdot, \cdot)$ over $\mathbb{R}^n \times \mathbb{R}^n$, alignment is defined as the expected distance between the normalized embeddings of positive pairs in contrastive learning. We use $f(\cdot)$ to indicate $L_2$ normalized representations. The alignment loss can be expressed using the following formula:
\begin{equation}
	\label{eq:alignment}
	\mathcal L_{\text{align}}(f; \alpha)\triangleq \underset{(x, x^+)\sim p_{\text{pos}}}{\mathbb{E}} \| f(x) - f(x^+) \|_2^\alpha, \quad \alpha > 0
\end{equation}

On the other hand, uniformity is the measurement of how uniformly the embeddings are distributed. Given the data distribution $p_\text{data}(\cdot)$ over $\mathbb{R}^n$ and the Gaussian potential kernel $G_t: \mathcal{S}^d \times \mathcal{S}^d \rightarrow \mathbb{R}_{+}$~\cite{cohn2007universally}, the uniformity loss is defined as the logarithm of the average pairwise Gaussian potential:
\begin{equation}
\label{eq:uniformity}
	\begin{aligned}
	\mathcal{L}_{\text {uniform }}(f ; t) & \triangleq \log \underset{\substack{x, y \sim p_{\text {data }}}}{\mathbb{E}}\left[G_t(f(x), f(y))\right] \\
	& =\log \underset{\substack{\text { i.i.d } \\
	x, y \sim p_{\text {data }}}}{\mathbb{E}}\left[e^{-t\|f(x)-f(y)\|_2^2}\right] \\
	& = -\log \underset{\substack{\text { i.i.d } \\
	x, y \sim p_{\text {data }}}}{\mathbb{E}}\left[e^{2t\left(1 - f(x)^\top f(y)\right)} \right], \quad t>0 \\
	\end{aligned}
\end{equation}

The alignment and uniformity metrics are well-suited to the central objective of contrastive learning: the embeddings of positive instances should stay close to each other, while the embeddings for random instances should scatter on the hypersphere. Furthermore, these two metrics are equally in strong alignment with collaborative filtering~\cite{wang2022towards}, where items of interest to users can form positive pairs with users. Here similarly, the normalized embeddings of positive user-item pairs should remain close, while the normalized embeddings of all users and items should be uniformly distributed on the hypersphere:
\begin{equation}
\label{eq:au1}
\begin{aligned}
\mathcal L_{\text{align}} &= \underset{(z_u, z_i)\sim p_{\text{pos}}}{\mathbb{E}} \| f(z_u) - f(z_i) \|_2^2 \\
\end{aligned}
\end{equation}
\begin{equation}
\label{eq:au2}
\begin{aligned}\mathcal{L}_{\text{uniform}} =& -\frac{1}{2} \log \underset{z_{u_x}, z_{u_y} \sim p_{\text{user}}}{\mathbb{E}} \left[ e^{2-2f(z_{u_x})^\top f(z_{u_y}) }\right] \\
& -\frac{1}{2} \log \underset{z_{i_x}, z_{i_y} \sim p_{\text{item}}}{\mathbb{E}} \left[ e^{2-2 f(z_{i_x})^\top f(z_{i_y})} \right]
\end{aligned}
\end{equation}

\noindent where $p_\text{pos}(\cdot, \cdot)$ denotes a distribution of positive pairs of user and item embeddings, $p_\text{user}(\cdot)$ denotes the distribution of user embeddings and $p_\text{item}(\cdot)$ denotes the distribution of item embeddings.

\section{Twin Graph Contrastive Learning}

\begin{figure*}[ht]
	\vspace{-0.5cm}
    \centering
    \includegraphics[width=0.8\textwidth]{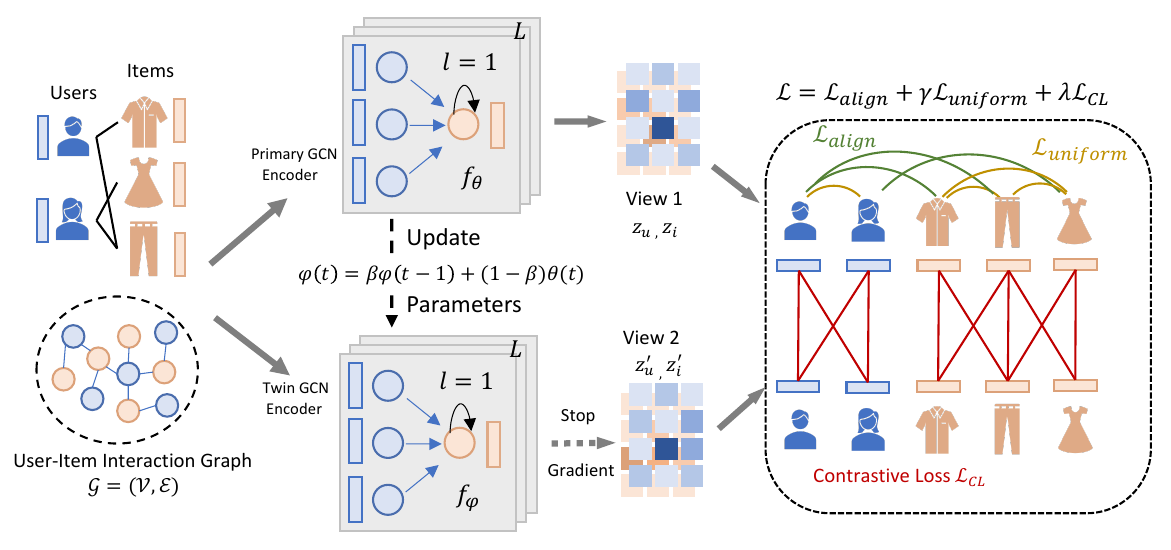}
    \caption{The overall framework of TwinCL.}
    \label{fig:framework}
\end{figure*}

In this section, we present the proposed Twin Graph Contrastive Learning (TwinCL) paradigm for graph collaborative filtering. The overall framework is illustrated in Figure~\ref{fig:framework}. TwinCL utilizes a twin encoder to generate effective contrastive views without augmentations on the graph structure or node representations. In addition, TwinCL leverages the properties of alignment and uniformity to measure the quality of representations, promoting the effectiveness of the twin encoder. Lastly, we also analyze the complexity of TwinCL to demonstrate its efficiency.

\subsection{Motivation}

Existing graph-based recommendation approaches often rely on random augmentations, which can introduce noise and disturb the structural and semantic information of the graph, especially when the graph is sparse with rare user-item interactions. Moreover, these approaches require trial-and-error selection of augmentation methods and parameters. We need to select an augmentation method with the appropriate magnitudes of perturbation, such as dropout rate~\cite{wujc2021self} or noise rate~\cite{yu2022graph, xia2022simgrace}. However, we notice that the optimal dropout rate or noise rate for augmentation is different depending on training epochs and iterations. In the initial stages of training, employing a higher dropout rate or perturbation leads to enhanced contrastive learning outcomes. However, as the model achieves a certain number of epochs, to mitigate overfitting and facilitate further improvements, it becomes desirable to reduce the dropout rate, thereby promoting greater similarity among augmentation views and achieving superior results. To mitigate the adverse effects of a fixed dropout rate, an alternative approach involves the use of a learnable neural generator~\cite{jiang2023adaptive} to create contrastive views. However, this method may not be efficient in collaborative filtering scenarios, as training an additional generator can be burdensome. We aim to identify an approach that is computationally efficient and allows for larger perturbations during the early stages of training, transitioning to milder perturbations in the later stages. 

In addition, the underutilization of alignment and uniformity optimization in GCL models represents an unexplored opportunity for improving recommendation performance. Most existing GCL methods~\cite{yu2022graph, lin2022improving, xia2022simgrace} merely analyze the alignment and uniformity of learned representations without leveraging these properties. Therefore, our research endeavors to design a GCL model that discards random augmentations in favor of directly optimizing alignment and uniformity, seeking to overcome data sparsity and enhance the robustness when subjected to significant perturbations during the early stages of training.

\subsection{Twin Encoder}

We introduce a Twin Encoder for contrastive learning. Our encoders consist of the primary graph encoder denoted as $f_\theta(\cdot)$ and its twin encoder counterpart, a momentum-updated encoder denoted as $f_\varphi(\cdot)$. The parameters of the primary encoder $f_\theta$ are represented as $\theta$, while the parameters of the twin encoder $f_\varphi$ are denoted as $\varphi$. The parameters refer to the learnable user and item embeddings within the encoders. These two encoders work in collaboration, with the twin encoder $f_\varphi$ dynamically adjusting its contrastive view generation in response to updates to the primary encoder $f_\theta$. This approach diverges from traditional methods involving random data augmentation or embedding perturbation, while preserving structural and semantic information, resulting in better representations.

\subsubsection{Twin Encoder Updating Mechanism}

The updating mechanism of the twin encoder can be represented by the following equation:
\begin{equation}
	\varphi(t)= \begin{cases} \theta(0) & t=0 \\ \beta \varphi(t-1)+ (1-\beta) \theta(t) & t>0 \end{cases}
\end{equation}

When training the model, the primary encoder's parameters $\theta$ are updated by back-propagation, while the twin encoder copies the parameters $\varphi$ from $\theta$ and updates itself ignoring its gradient. For iteration $n$, the parameters of the twin encoder $\varphi(n)$ are:
\begin{equation*}
\begin{aligned}
	\varphi(n) &= \beta \varphi(n-1) + (1-\beta)\theta(n-1) \\
	&= \beta(\beta \varphi(n-2) + (1-\beta) \theta(n-2)) + (1-\beta)\theta(n-1) \\
	& \cdots \\
	&= \beta^{n-1} \theta(0) + (1-\beta) \sum_{t=1}^n \beta^{n-t} \theta(t-1) \\
\end{aligned}
\end{equation*}
The momentum coefficient $\beta \in (0, 1)$ represents the weight or influence of the historical parameter values of the graph encoder $f_\theta$ when updating the parameters of the twin encoder $f_\varphi$ during training. It plays a crucial role in controlling the impact of past updates on the current update of the twin encoder's parameters. The momentum-based updating mechanism ensures a balance between historical parameter values and current gradients. If we use a larger $\beta$, it means that a larger portion of the previous parameters of $f_\varphi$ is retained, resulting in smoother updates and faster convergence in some cases. In the experiments, $\beta = 0.9$ is usually a common choice with an appropriate evolving speed for an ideal contrastive learning effect on various datasets.

Our twin encoder updating mechanism guarantees diversity between the twin encoders and the generated contrastive views in the early stages of training. As the model approaches convergence, the similarity between the twin encoders increases significantly. This feature enhances the performance of contrastive learning compared to fixed-rate perturbation methods, while also ensuring efficient generation of contrastive views. %

TwinCL is purpose-built for collaborative filtering, ensuring better performance and efficiency, setting it apart from existing models featuring dual encoders such as MERIT~\cite{jin2021multi} and SimGRACE~\cite{xia2022simgrace} for the following reasons: (1) TwinCL does not require data augmentations, preserving both the structural and semantic information of graphs, while MERIT relies on graph augmentations. (2) SimGRACE employs a less efficient perturbation method for collaborative filtering, and its emphasis on graph-level representations renders it less effective for filtering tasks compared to TwinCL.

The graph encoder $f_\theta$ and its twin encoder $f_\phi$ extract node-level representations $z_u, z_i$ and $z_u', z_i'$ for contrastive learning, respectively, through graph convolution without feature transformation and nonlinear activation, following the propagation rule of LightGCN:
\begin{equation} 
	z_u^{(l)} = \text{GraphConv}(z_u^{(l-1)},  \{z_i^{(l-1)}: i \in \mathcal N_u\}) = \tilde{\mathbf{A}}^i z_u^{(l-1)}
\end{equation}
\begin{equation}
z_u = \frac{1}{L+1} \sum_{i=0}^L z_u^{(i)}  = \frac{1}{L+1} \sum_{i=0}^L \tilde{\mathbf{A}}^i z_u^{(0)}	
\end{equation}
where $\tilde{\mathrm{A}} \in \mathcal{R}^{(|\mathcal{U}| + |\mathcal{I}| ) \times (|\mathcal{U}| + |\mathcal{I}| ) }$ is the symmetrically normalized adjacency matrix of the user-item graph, and $L$ is the number of layers. We use $z_u$ and  $z_i$ for the final prediction $s(u,i) = z_u^\top z_i$, which is the ranking score for recommendation.

\subsubsection{Contrastive Learning between the Twins}

The contrastive learning between the primary encoder and the twin encoder is to maximize the similarity between the representations of positive pairs generated by the twins $f_\theta$ and $f_\varphi$ while simultaneously encouraging uniformity in the embedding space. This is accomplished through the InfoNCE~\cite{oord2018representation} loss formulation:
\begin{equation}
\mathcal{L}_\mathrm{CL}=\sum_{i \in \mathcal{B}}-\log \frac{\exp(z_i^\top z_i^\prime /\tau)}{\sum_{j \in \mathcal{B}} \exp \left(z_i^\top z_j^\prime / \tau\right)}
\end{equation}
In this equation, $z_i$ and $z_i^\prime$ are the representations of user $i$ or item $i$ generated by the graph encoder $f_\theta$ and its twin encoder $f_\varphi$. $\tau$ is the temperature parameter that scales the logits before applying the softmax. $\mathcal{B}$ is the batch of embeddings.

For alignment, we desire positive pairs to have similar representations. The term $\exp(z_i^\top z_i^\prime /\tau)$ in the numerator measures the similarity between positive pairs. A high value indicates effective alignment between the two views of data point $i$. We aim to align the representations of positive pairs generated by twin encoders. During the initial stages of training, the similarity between twin encoders is relatively low because of the updating mechanism, resulting in smaller similarities of positive pairs. Aligning such positive pairs facilitates accelerated contrastive learning.

Furthermore, uniformity ensures that representations are evenly distributed in the embedding space. The term $\sum_{j \in \mathcal{B}} \exp \left(z_i^\top z_j^\prime / \tau\right)$ in the denominator sums over similarities of the current data point $i$ with all other data points in the batch. When representations are uniform, each term in this summation would be small, as there would not be any other representation too similar to $z_i$. Minimizing the InfoNCE loss would push the model to ensure that while positive pairs are aligned, other pairs are not too similar, thereby promoting uniformity.

\subsection{Alignment and Uniformity Optimization}

We delve into the specifics of how we adapt the optimization of alignment and uniformity for our collaborative filtering model. In conventional recommendation systems, the Bayesian Personalized Ranking (BPR)~\cite{rendle2012bpr} loss is a popular choice for optimizing the ranking of items. However, it comes with its own set of limitations, primarily due to its sensitivity to negative sampling~ and the difficulty in capturing complex user-item interactions. The performance of negative sampling may be compromised by the low quality of sampled negative examples and slow convergence. To overcome these limitations, our model employs two main optimization losses. The \textit{Alignment Loss} ensures that embeddings of similar users and items in the graph are closer in the representation space, leading to more accurate recommendations. The \textit{Uniformity Loss} ensures that the embeddings are uniformly distributed over the representation space, reducing the chance of the model overfitting to a particular item or user category, and making the model more robust and generalizable.

The synergy of alignment and uniformity losses allows our model to achieve commendable performance right from the onset. We achieve good performance from the beginning epochs, which helps TwinCL achieve better performance through contrastive learning. A previous study~\cite{wang2022towards} has shown that given the existence of perfectly aligned and uniform encoders, they form the exact minimizers of the BPR loss $\mathcal L_{\text{BPR}}$:
\begin{equation}
\label{eq:bpr}
\begin{aligned}
\mathcal{L}_{\text{BPR}} &=\underset{(u, i) \sim p_{\text{pos}}}{\mathbb{E}}-\log \left[\operatorname{sigmoid}\left(s(u, i)-s\left(u, i^{-}\right)\right)\right] \\
& =\underset{(u, i) \sim p_{\text{pos}}}{\mathbb{E}} - \log \left(\frac{\exp(z_u^\intercal z_i)}{\exp(z_u^\intercal z_i) + \exp(z_u^\intercal z_i^-)} \right) \\
& =\underset{(u, i) \sim p_{\text{pos}}}{\mathbb{E}} - z_u^\intercal z_i + \log \left(\exp(z_u^\intercal z_i)+\exp(z_u^\intercal z_i^-)\right) \\
& \geq \underset{(u, i) \sim p_{\text{pos}}}{\mathbb{E}} -1 + \log \left(e+\exp(z_u^\intercal z_i^-)\right) \\
\end{aligned}
\end{equation}
When the representations of positive pairs of user and item have perfect alignment and all representations are uniformly distributed on the hypersphere, $\mathcal L_\text{BPR}$ is minimized. Therefore, the alignment and uniformity losses are able to substitute the BPR loss to optimize the recommendation models. By directly optimizing alignment loss $\mathcal L_\text{align}$ and uniformity loss $\mathcal L_\text{uniform}$, the models inherently align the embeddings of positive user-item pairs as well as ensure their uniform distribution, leading to a richer representation that outperforms traditional methods reliant on BPR loss.

As Equation~\ref{eq:au1} and ~\ref{eq:au2} show the losses of alignment and uniformity, we can use these two objectives as the substitute for $\mathcal L_\text{BPR}$. It becomes more straightforward and does not require negative sampling for negative item embeddings, impelling easier optimization as well as faster convergence.

Most previous GCL models are typically designed to target the BPR loss instead of the perspective of alignment and uniformity. Therefore, directly transplanting methods that optimize alignment and uniformity onto GCL models may not yield satisfactory results. TwinCL, on the other hand, aligns well with these key properties. The effectiveness of the twin encoder in contrastive learning depends on its parameters, and alignment and uniformity can ensure that the twin encoder obtains higher-quality parameters from the primary encoder in the early epochs of training, thereby facilitating better contrastive signals between the two encoders.

\subsection{Training Strategy of TwinCL}

The training strategy of TwinCL is to jointly optimize alignment loss, uniformity loss, and contrastive loss:
\begin{equation}
	\mathcal L = \mathcal L_\text{align} + \gamma \mathcal  L_\text{uniform} + \lambda \mathcal L_\text{CL} + \lambda_\text{r} \|\Theta\|_2^2
\end{equation}
where $\gamma$ is the weight of uniformity loss and $\lambda$ is the weight of contrastive loss, respectively. $\lambda_\text{r}$ is the weight of $L_2$ regularization loss, and $\Theta$ is the set of model parameters in $\mathcal L_\text{align}$ since $\mathcal L_\text{uniform}$ and $\mathcal L_\text{CL}$ do not introduce additional parameters. The training process is described in Algorithm~\ref{alg:TwinCL}.

\algnewcommand{\Initialize}[1]{%
	\State \textbf{Initialize:}
	\Statex \hspace*{\algorithmicindent}\parbox[t]{.8\linewidth}{\raggedright #1}
}

\begin{algorithm}[htb]
	\setlength{\baselineskip}{13bp}
	\caption{Twin Graph Contrastive Learning (TwinCL)}\label{alg:TwinCL} 
	\begin{algorithmic}[1]
		\State\textbf{Input:} bipartite graph $\mathcal G = \{\mathcal U \cup \mathcal I, \mathcal E\}$; training set $\mathcal X$; weight of uniformity loss $\gamma$; weight of contrastive learning loss $
		\lambda$.
		\Initialize{ 
        $\theta \leftarrow$ Initialized encoder parameters \\
		$\varphi \leftarrow \theta$ \\
		}
	
	\While {\textit{not convergence}}
		\For {$x$ in Dataloader($\mathcal X$)}
			\State $z_u, z_i$ $\leftarrow$ Graph convolution by $f_\theta(\mathcal G)$
			\State Calculate $\mathcal L_\text{align}(z_u, z_i)$ and $\mathcal  L_\text{uniform}(z_u, z_i)$	
			\State $z_u^\prime, z_i^\prime \leftarrow$ Graph convolution by $f_\varphi(\mathcal G)$
			\State Calculate contrastive loss $\mathcal L_\text{CL}(z_u, z_u^\prime, z_i, z_i^\prime)$
			\State $\mathcal L \leftarrow \mathcal L_\text{align} + \gamma \mathcal  L_\text{uniform} + \lambda \mathcal L_\text{CL}$ 
			\State Update $\theta$ by back-propagation, minimizing $\mathcal L$
			\State $\varphi \leftarrow \beta \varphi + (1 - \beta) \theta$
			\EndFor
		\EndWhile
		\State $z_u, z_i$ $\leftarrow$ Graph convolution by $f_\theta(\mathcal G)$
		\State\textbf{Return:} Final user embeddings $z_u$ and item embeddings $z_i$
	\end{algorithmic}
\end{algorithm}

\subsection{Time Complexity Analysis}

We analyze the time complexity of TwinCL and compare it with other GCL models. All models use LightGCN as the graph encoder for a fair comparison. Let $|\mathcal E|$ be the edge number in the graph, $d$ be the embedding size, $B$ denote the batch size, $M$ represent the node number in a batch, and $\rho$ denote the edge keep rate in SGL. When training the model, the graph convolution operations dominate computation time. For a LightGCN encoder with $L$ layers, the adjacency matrix has $2|\mathcal E|$ non-zero elements, so each graph convolution takes $O(2|\mathcal E|d)$. Both SGL and SimGCL need three times as many graph convolutions as they generate two additional contrastive views. The graph convolution costs per epoch for SGL and SimGCL are $O((2 + 4\rho)|\mathcal E|Ld)$ and $O(6|\mathcal E|Ld)$ respectively. In comparison, TwinCL which produces only a single view with a twin encoder, has a complexity of $O(4|\mathcal E|Ld)$, explaining its faster training compared to SGL and SimGCL.

\section{Experiments}

\subsection{Experimental Settings}

\subsubsection{Datasets}

We conduct experiments on three public benchmark datasets in real-world e-commerce scenarios: Yelp2018~\cite{wang2019neural}, Amazon-Book~\cite{he2016ups}, and Alibaba-iFashion~\cite{chen2019pog}. 

\begin{table}[ht]
\normalsize
  \centering
  \caption{Statistics of the experimented datasets.}
  \label{tab:dataset}
  \resizebox{0.47\textwidth}{!}{%
  \begin{tabular}{lcccc}
    \toprule
    \multicolumn{1}{l}{\textbf{Dataset}} & \multicolumn{1}{c}{\textbf{\# Users}} & \multicolumn{1}{c}{\textbf{\# Items}} & \multicolumn{1}{c}{\textbf{\# Interactions}} & \multicolumn{1}{c}{\textbf{Density}} \\ 
    \midrule
    Yelp2018 & 31,688 & 38,048 & 1,561,406 & 0.130\%\\
    Amazon-Book & 52,643 & 91,599 & 2,984,108 & 0.062\%\\
    Alibaba-iFashion & 300,000 & 81,614 & 1,607,813 & 0.007\%\\
  \bottomrule
\end{tabular}}
\end{table}

\subsubsection{Baselines}

\begin{table*}[ht!]
\renewcommand{\arraystretch}{0.8}
\tabcolsep=6.5pt
\centering
\caption{Top-\textit{K} recommendation performance comparison on three datasets. The best results are highlighted in bold, and the best baselines are underscored. The relative improvements in comparison to the best baselines are indicated as Improv.}
\begin{tabular}{clccccccclc}
    \toprule
    \multicolumn{2}{c}{Setting} & \multicolumn{7}{c}{Baselines} & \multicolumn{2}{c}{Ours}\cr 
    \cmidrule(lr){1-2} \cmidrule(lr){3-9} \cmidrule(lr){10-11}
    Dataset & \multicolumn{1}{c}{Metric} & BPRMF & LightGCN & BUIR & DirectAU & NCL & SGL & SimGCL & TwinCL & Improv. \cr
    \midrule
    \multirow{7}{*}{\rotatebox{90}{Yelp2018}} 
    & Recall@10 & 0.0283 & 0.0350 & 0.0291 & 0.0408 & 0.0394 & 0.0396 & \underline{0.0425} & \textbf{0.0440} & 3.53\% \cr
    & Recall@20 & 0.0498 & 0.0639 & 0.0497 & 0.0692 & 0.0673 & 0.0677 & \underline{0.0719} & \textbf{0.0736} & 2.36\% \cr
    & Recall@50 & 0.0988 & 0.1158 & 0.0976 & 0.1326 & 0.1293 & 0.1302 & \underline{0.1358} & \textbf{0.1385} & 1.99\% \cr
    \cmidrule(lr){2-2} \cmidrule(lr){3-9} \cmidrule(lr){10-11}
    & NDCG@10 & 0.0321 & 0.0399 & 0.0330 & 0.0472 & 0.0450 & 0.0451 & \underline{0.0485} & \textbf{0.0505} & 4.12\% \cr
    & NDCG@20 & 0.0401 & 0.0525 & 0.0404 & 0.0573 & 0.0553 & 0.0555 & \underline{0.0593} & \textbf{0.0611} & 3.04\% \cr
    & NDCG@50 & 0.0581 & 0.0697 & 0.0582 & 0.0808 & 0.0782 & 0.0785 & \underline{0.0827} & \textbf{0.0851} & 2.90\% \cr
    \midrule
    \multirow{7}{*}{\rotatebox{90}{Amazon-Book}} 
    & Recall@10 & 0.0172 & 0.0225 & 0.0157 & 0.0250 & 0.0259 & 0.0275 & \underline{0.0306} & \textbf{0.0321} & 4.90\% \cr
    & Recall@20 & 0.0303 & 0.0411 & 0.0271 & 0.0426 & 0.0442 & 0.0478 & \underline{0.0510} & \textbf{0.0534} & 4.71\% \cr
    & Recall@50 & 0.0605 & 0.0746 & 0.0536 & 0.0829 & 0.0845 & 0.0881 & \underline{0.0915} & \textbf{0.0978} & 6.89\% \cr
    \cmidrule(lr){2-2} \cmidrule(lr){3-9} \cmidrule(lr){10-11}
    & NDCG@10 & 0.0183 & 0.0239 & 0.0175 & 0.0279 & 0.0271 & 0.0293 & \underline{0.0324} & \textbf{0.0348} & 7.41\% \cr
    & NDCG@20 & 0.0235 & 0.0315 & 0.0217 & 0.0344 & 0.0343 & 0.0379 & \underline{0.0406} & \textbf{0.0430} & 5.91\% \cr
    & NDCG@50 & 0.0348 & 0.0436 & 0.0314 & 0.0491 & 0.0492 & 0.0521 & \underline{0.0553} & \textbf{0.0592} & 7.05\% \cr
    \midrule
    \multirow{7}{*}{\rotatebox{90}{iFashion}} 
    & Recall@10 & 0.0382 & 0.0603 & 0.0548 & 0.0728 & 0.0617 & 0.0761 & \underline{0.0778} & \textbf{0.0815} & 4.76\% \cr
    & Recall@20 & 0.0583 & 0.0906 & 0.0842 & 0.1086 & 0.0914 & 0.1109 & \underline{0.1138} & \textbf{0.1185} & 4.13\% \cr
    & Recall@50 & 0.0978 & 0.1470 & 0.1411 & 0.1769 & 0.1476 & 0.1743 & \underline{0.1781} & \textbf{0.1866} & 4.77\% \cr
    \cmidrule(lr){2-2} \cmidrule(lr){3-9} \cmidrule(lr){10-11}
    & NDCG@10 & 0.0214 & 0.0342 & 0.0305 & 0.0412 & 0.0350 & 0.0437 & \underline{0.0447} & \textbf{0.0471} & 5.37\% \cr
    & NDCG@20 & 0.0267 & 0.0422 & 0.0383 & 0.0506 & 0.0428 & 0.0529 & \underline{0.0543} & \textbf{0.0568} & 4.60\% \cr
    & NDCG@50 & 0.0348 & 0.0539 & 0.0501 & 0.0648 & 0.0543 & 0.0659 & \underline{0.0679} & \textbf{0.0708} & 4.27\% \cr
    \bottomrule
\end{tabular}
\label{tab:overall}
\end{table*}

We compare the performance of TwinCL with several baseline collaborative filtering and graph contrastive learning methods:
\begin{itemize}
[leftmargin=*,noitemsep,topsep=1.5pt]
    \item \textbf{BPRMF}~\cite{rendle2012bpr}: This algorithm employs a Bayesian approach to optimize matrix factorization to provide personalized ranking predictions.
    \item \textbf{LightGCN}~\cite{he2020lightgcn}: This is a simplified graph convolutional network emphasizing neighborhood aggregation to learn user and item embeddings.
    \item \textbf{BUIR}~\cite{lee2021bootstrapping}: This is a collaborative filtering method that does not require negative sampling to learn user and item embeddings.
    \item \textbf{DirectAU}~\cite{wang2022towards}: This is a method that discards the BPR loss and optimizes representation quality by directly enhancing alignment and uniformity on the hypersphere. We use LightGCN as its encoder in our experiments.
    \item \textbf{SGL}~\cite{wujc2021self}: This graph contrastive learning method performs data augmentations on the graph structure, such as node or edge dropout. We use edge dropout for data augmentation in our experiments.
    \item \textbf{NCL}~\cite{lin2022improving}: This approach uses a prototypical contrastive objective to leverage both structural and semantic neighbor relations.
    \item \textbf{SimGCL}~\cite{yu2022graph}: This method introduces noise to the embedding space to create contrastive views of user and item embeddings to enhance recommendation performance and training efficiency.
\end{itemize}

\subsubsection{Hyperparameter Settings}

For a fair comparison, all models are trained from scratch. All the embedding sizes are fixed to 64 and the embedding parameters are initialized with the Xavier method~\cite{glorot2010understanding}. We refer to the best hyperparameter settings reported in the original papers of baselines. The models are optimized with Adam optimizer~\cite{kingma2014adam} with a learning rate of 0.001 and a default mini-batch size of 2048. For all contrastive learning models, we empirically set the default temperature $\tau = 0.2$, which is the best value in most cases. The $L_2$ regularization coefficient is fixed at 0.0001. In regards to the tuning ranges for $\lambda$, $m$, and $\gamma$ and the sensitivity to hyperparameters in TwinCL, we provide more details and results in the appendix.

\subsection{Performance Comparison}
The evaluation metrics are Recall@\textit{K} and NDCG@\textit{K} computed by the all-ranking protocol -- all items that have not interacted with a user are the candidates. We choose 10, 20, and 50 as the value of \textit{K}.

Table~\ref{tab:overall} presents the overall performance of TwinCL and the seven baselines. From the results, we have the following observations and conclusions.

\begin{figure*}[ht]
    \centering
    \includegraphics[width=0.85\textwidth]{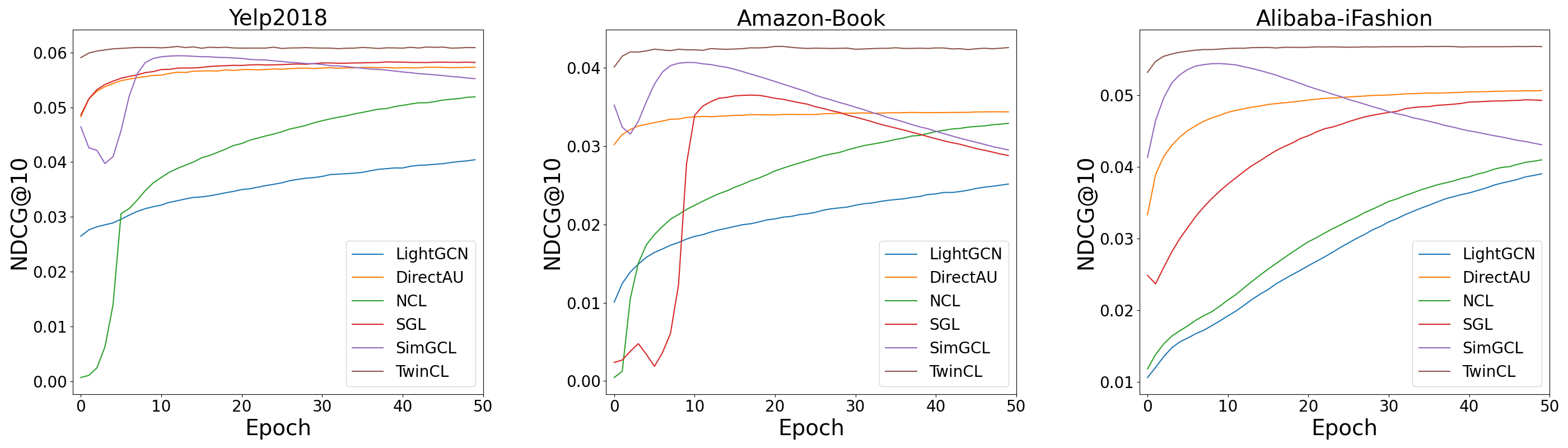}
    \caption{The learning curves in the first 50 epochs of baselines and TwinCL on three benchmark datasets.}
    \label{fig:learning_curve}
\end{figure*}

\begin{itemize}
[leftmargin=*,noitemsep,topsep=1.5pt]
	\item DirectAU and all GCL models improve the performance based on LightGCN, which shows that both contrastive learning and optimization of alignment and uniformity are effective in improving performance and tackling the data sparsity issue. 
	\item TwinCL achieves the best performance in all cases, which demonstrates that our simple method is effective in enhancing recommendations. The performance has significant improvement, especially on the two sparser datasets -- Amazon-Book and Alibaba-iFashion.
	\item Complex data augmentation and contrastive learning methods do not necessarily yield better results, as seen in examples like NCL and SGL, which also exhibit poorer running time performance. In contrast, TwinCL employs straightforward techniques and achieves better results.
\end{itemize}

\begin{figure}[ht]
    \centering
    \includegraphics[width=0.45\textwidth]{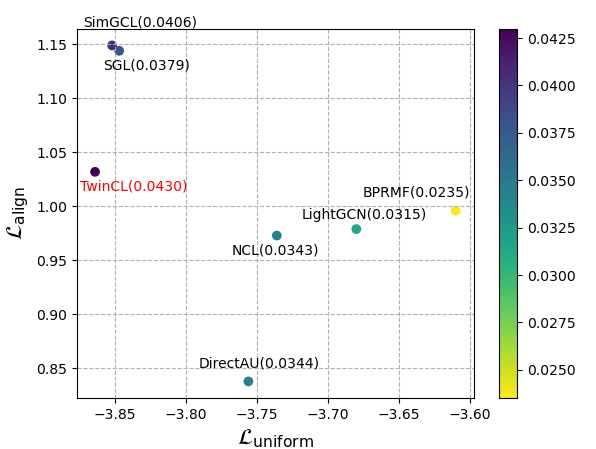} 
    \caption{$\mathcal L_\text{align}-\mathcal L_\text{uniform}$ plot of different models on Amazon-Book. For $\mathcal L_\text{align}$ and $\mathcal L_\text{uniform}$, lower values are better. Colors and numbers in parentheses indicate NDCG@20.}
    \label{fig:lau}
\end{figure}

In Figure~\ref{fig:lau} we show the alignment and uniformity of baselines and TwinCL. Compared to SGL and SimGCL, TwinCL effectively reduces alignment loss while preserving a low level of uniformity loss. This achievement is attributed to TwinCL's direct optimization of alignment and uniformity. TwinCL manages to achieve strong recommendation performance by maintaining a low level of alignment and uniformity losses.

\subsection{Ablation and Effectiveness Analyses}

We conduct ablation studies on TwinCL by showing how alignment and uniformity losses and the twin encoder affect its training and performance. We also conduct experiments on modified versions of SGL, SimGCL, and TwinCL. Specifically, SGL-AU and SimGCL-AU incorporate alignment and uniformity optimization. TwinCL-BPR, on the other hand, is optimized using the BPR loss, without the inclusion of alignment and uniformity losses. The results of these variants are presented in Table~\ref{tab:ablation}. The combination of the BPR loss and TwinCL does not yield optimal performance because the BPR loss leads to slower convergence and the parameters of the twin encoder cannot receive effective updates.

\begin{table}[ht]
\renewcommand\arraystretch{0.9}
	\caption{Performance comparison of GCL model variants.}
	\small
	\label{tab:ablation}
	\begin{center}
	\resizebox{0.47\textwidth}{!}{
	\begin{tabular}{l|cc|cc}
		\toprule
		\multirow{2}{*}{\textbf{Method}}&\multicolumn{2}{c}{\textbf{Yelp2018}}& \multicolumn{2}{c}{\textbf{Amazon-Book}} \cr
		\cmidrule(lr){2-3}\cmidrule(lr){4-5}&\textbf{Recall@20} & \textbf{NDCG@20}  & \textbf{Recall@20} & \textbf{NDCG@20} \\ \hline
		SGL  & 0.0677 & 0.0555 & 0.0478 & 0.0379	\\
		SGL-AU  & 0.0701 & 0.0582 & 0.0493 & 0.0395  \\	
		\hline
		SimGCL  & 0.0719 & 0.0593 & 0.0510 & 0.0406  \\
		SimGCL-AU  & 0.0723 & 0.0599 & 0.0516 & 0.0412  \\	
		\hline
		TwinCL-BPR  & \underline{0.0724} & \underline{0.0603} & \underline{0.0518}  & \underline{0.0415}  \\	
		TwinCL  & \textbf{0.0736} & \textbf{0.0611} & \textbf{0.0534} & \textbf{0.0430}  \\		
		\bottomrule
	\end{tabular}}
	\end{center}
\end{table}

\noindent \textbf{Impact of Alignment and Uniformity.} As observed in Figure~\ref{fig:learning_curve}, optimizing alignment and uniformity in TwinCL leads to faster convergence, improved training stability, and reduced overfitting with increasing training epochs. Table~\ref{tab:ablation} demonstrates that this optimization approach not only benefits TwinCL but also other GCL recommendation models, though the improvement varies across different GCL models.

\noindent \textbf{Impact of Twin Encoder.} The twin encoder can deliver desired and fast contrastive learning outcomes in the experiments. It simplifies the generation of contrastive views, resulting in reduced graph convolution operations and shorter training times. When prioritizing alignment and uniformity optimization, the twin encoder demonstrates higher improvements compared to other GCL models in Table~\ref{tab:ablation}. Leveraging the promising initial performance gains achieved through alignment and uniformity optimization in the early training epochs, the twin encoder updates parameters of superior quality. This enhancement positively impacts contrastive learning with the primary encoder, leading to faster training convergence and better overall performance. In comparison, the SimGCL encoder with random noise perturbation does not effectively leverage the improvements in contrastive learning brought about by alignment and uniformity optimization.

\subsection{Training Speed Comparison}

\noindent \textbf{Convergence Speed Comparison.} TwinCL enjoys faster convergence due to the optimization of alignment and uniformity as well as the effective contrastive learning by the twin encoder. The incorporation of alignment and uniformity losses, coupled with contrastive learning, enhances the model's ability to acquire representations of superior quality. We can observe that TwinCL exhibits a fast convergence speed. In Figure~\ref{fig:learning_curve}, TwinCL and SimGCL demonstrate similar convergence rates, converging in approximately 10-20 epochs across three datasets. In contrast, SGL requires 18-40 epochs to converge on different datasets, while NCL requires 70-100 epochs for convergence.

\noindent \textbf{Running Time Comparison.} In Table~\ref{tab:time}, we present a comparative analysis of the running time per epoch for different GCL models. All experiments are conducted on a single Nvidia A5000. Notably, TwinCL exhibits a shorter running time per epoch when compared with SGL and SimGCL. This reduced computational demand is a direct consequence of TwinCL's efficiency in graph convolution operations. Furthermore, when taking into account the convergence speed, TwinCL is the most time-efficient option for training and achieving desirable performance among these GCL models.

\begin{table}[ht]
\renewcommand\arraystretch{0.9}
\normalsize
\tabcolsep=9.0pt
\caption{Running time per epoch of GCL models}
\begin{tabular}{l|c|c|c}
\hline
Method & Yelp2018 & Amazon-Book & iFashion \\
\hline
\hline
NCL  & 74.3s & 207.4s & 143.6s      \\
\hline
SGL & 106.8s & 330.5s &  148.9s \\
\hline
SimGCL &  112.1s  &  349.1s & 170.2s  \\
\hline     
TwinCL & 89.4s & 238.9s & 115.1s \\
\hline
\end{tabular}
\label{tab:time}
\vspace{-0.4cm}
\end{table}

\subsection{Effectiveness in Mitigating Popularity Bias}

\begin{figure}[ht]
	\centering
	\includegraphics[width=.45\textwidth]{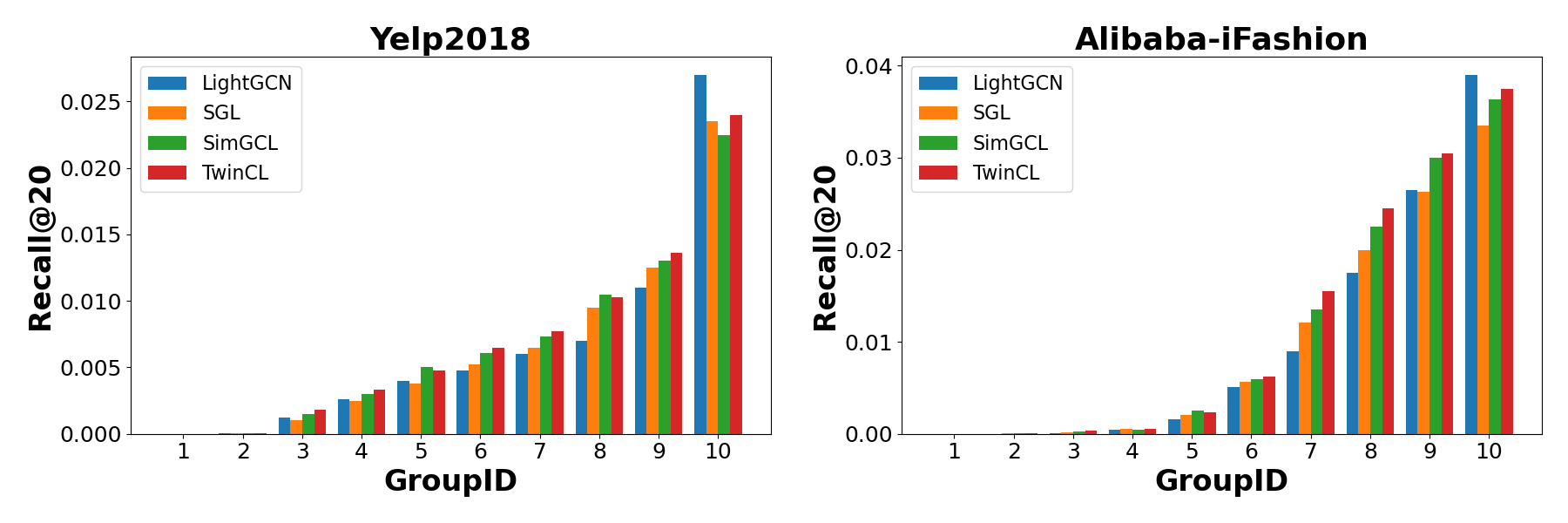}
	\caption{Ability to mitigate popularity bias.}
	\label{fig:bias}
\end{figure}

To address the issue of popularity bias~\cite{chen2023bias, zhu2021popularity}, the representations of both users and items are expected to be distributed more evenly, which can be accomplished by optimizing both uniformity loss and contrastive loss. 
We split items into ten groups based on popularity, with each group containing an equal number of interactions. The groups with higher IDs included more popular items. We conduct our experiments on the Yelp2018 and Alibaba-iFashion datasets to evaluate the ability to tackle popularity bias and data sparsity, and the results are presented in Figure ~\ref{fig:bias}. 
We can observe that the improvements achieved by TwinCL are mainly from items with lower popularity. Note that LightGCN tends to prioritize popular items and achieves higher recall on Group 10. However, LightGCN's recommendations are inaccurate, and its NDCG is also lower compared to other models. In this regard, TwinCL outperforms both LightGCN and other GCL models, and its exceptional performance in long-tail items aligns with users' real needs.

\subsection{Comparison between Twin Encoders}

\begin{figure}[ht]
	\centering
	\includegraphics[width=.45\textwidth]{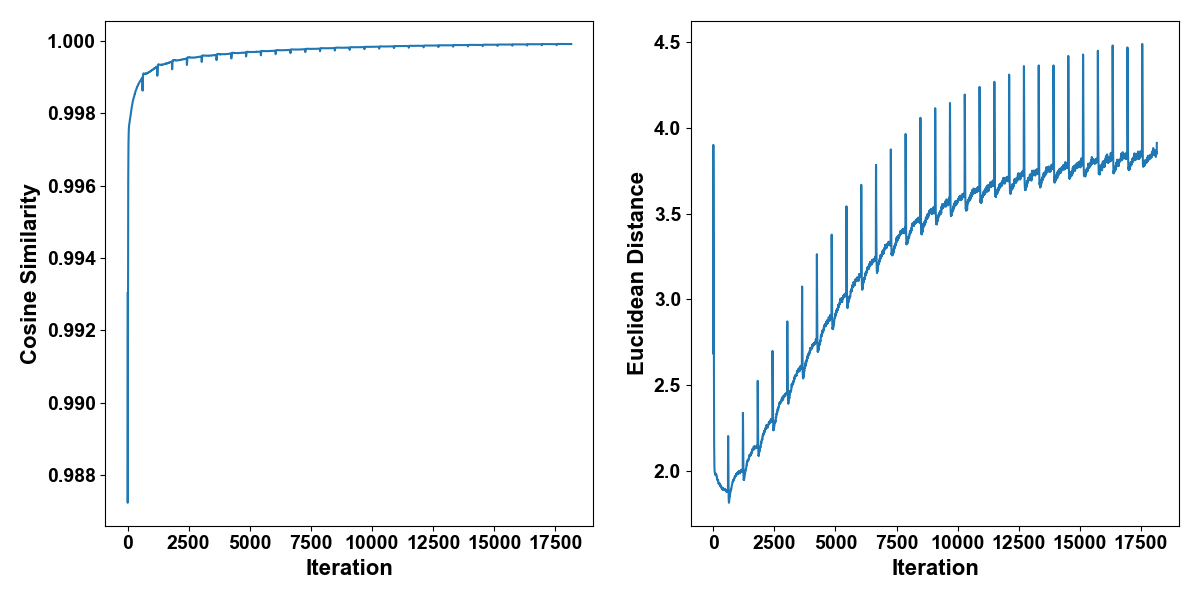}
	\caption{Similarity and distance curves of twin encoders}
	\label{fig:curve}
\end{figure}

In this section, we evaluate the similarity and difference between the twin encoders using both the cosine similarity and the Euclidean distance of their parameters. Specifically, the parameters are the node embeddings within the backbone encoder of LightGCN. The corresponding curves for cosine similarity and Euclidean distance while training TwinCL on the Yelp2018 dataset are in Figure ~\ref{fig:curve}. The number of iterations depends on the combination of epochs and batch size. In our setting, we have 763 iterations per epoch.

We can observe that in the initial 2500 iterations, particularly within the initial three epochs, there is a significant increase in cosine similarity. This leads to the creation of more diverse contrastive views by the twin encoders. In subsequent iterations, the gradient of the cosine similarity curve gradually decreases, ultimately stabilizing as the convergence approaches. Consequently, the generated views begin to exhibit greater similarity to each other. In terms of the Euclidean distance, it experiences a rapid decline within the first 763 iterations, corresponding to the first epoch, followed by a gradual increase with a diminishing gradient. The increasing distance can be attributed to the progressively enriched node embeddings over time, particularly evident in the embeddings of positive pairs. Note that fluctuations in the curves are from the start of a new training epoch. Hence, the cosine similarity and Euclidean distance curves are aligned with the design motivation behind our momentum-updating mechanism. 

\subsection{Hyperparameter Sensitivity Analysis}

\begin{figure}[ht]
	\centering
	\includegraphics[width=.45\textwidth]{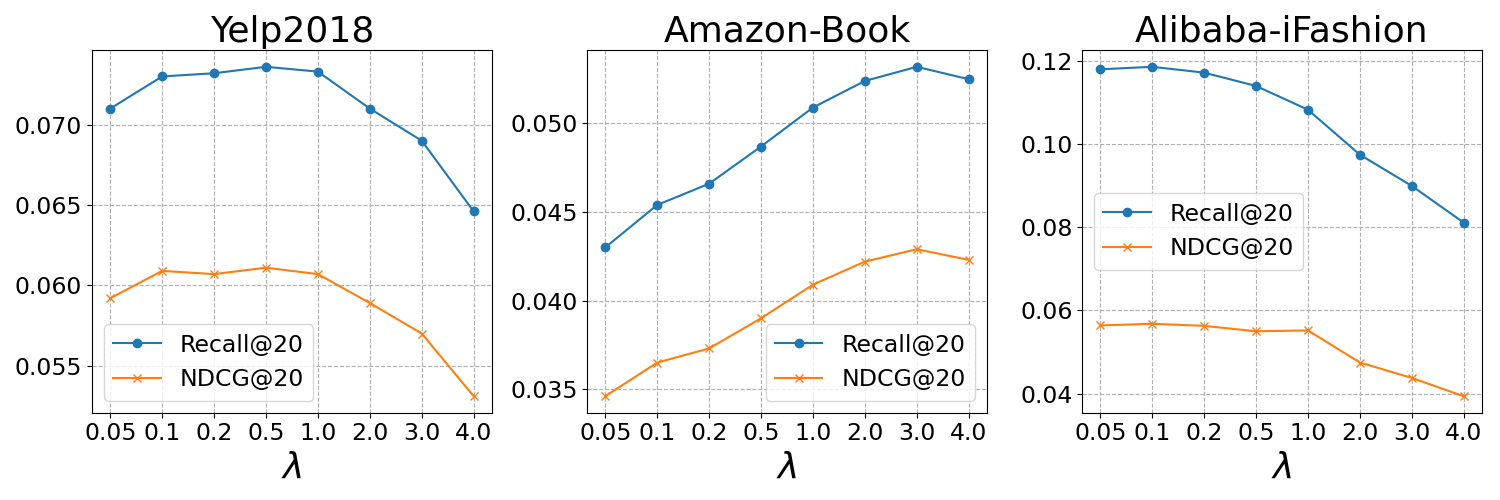}
	\caption{Influence of the magnitude $\lambda$ of TwinCL. }
	\label{fig:lambda}
\end{figure}

\textbf{Impact of Coefficient $\lambda$.} Our model's performance is sensitive to the contrastive learning loss weight $\lambda$. As we can observe in Figure~\ref{fig:lambda}, for Yelp2018, Amazon-Book, and Alibaba-Ifashion, the optimal lambda values for TwinCL are 0.5, 3.0, and 0.1, respectively. Without selecting an appropriate lambda, the model's performance exhibits a significant decrease. When $\lambda$ is too high, the model emphasizes close grouping of similar items at the expense of distinguishing dissimilar ones. Conversely, a too-low $\lambda$ prioritizes separating dissimilar items, potentially hampering the effective grouping of similar items.

\begin{figure}[ht]
	\centering
	\includegraphics[width=.45\textwidth]{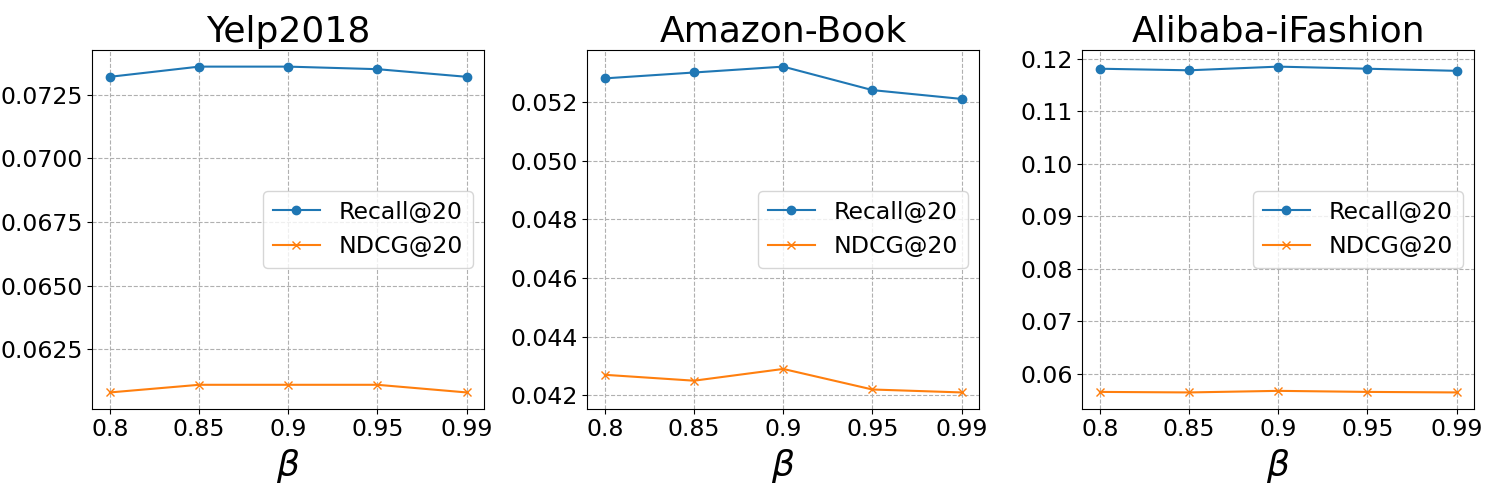}
	\caption{Influence of the magnitude $\beta$ of TwinCL. }
	\label{fig:beta}
\end{figure}

\noindent \textbf{Impact of Coefficient $\beta$.} The momentum coefficient $\beta$ controls the evolving speed of the twin encoder and influences the similarity of parameters between the twin encoder and the primary encoder. Compared to $\lambda$, the impact of momentum coefficient $\beta$ on the model's performance is relatively minor when it ranges from 0.8 to 0.99. The coefficient $\beta \in [0.8, 0.99]$ consistently yields favorable experimental results. As shown in Figure~\ref{fig:beta}, setting $\beta$ to 0.9 delivers excellent performance across all three datasets. The performance of our model on the above three datasets is not greatly affected by $\beta$. This suggests that our model exhibits robustness in relation to the hyperparameter $\beta$ selection, in contrast to prior methods like SGL and SimGCl, which are notably sensitive to variations in dropout rate or noise perturbation magnitude.

\noindent \textbf{Impact of Coefficient $\gamma$.} The weight of uniformity loss $\gamma$ controls the desired degree of uniformity. We have experimented with different values of $\gamma$ and observed that compared to the contrastive learning loss weight $\lambda$, the overall performance of the model is less sensitive to the value of $\gamma$. In other words, when selecting appropriate contrastive learning $\lambda$ and $\beta$ values, using different $\gamma$ values can still achieve relatively decent experimental results. Contrastive learning can mitigate the impact of $\gamma$, i.e. the weight of uniformity loss, on overall performance, because the InfoNCE loss can optimize uniformity as well. For the Yelp2018, Amazon-Book, and Alibaba-iFashion datasets, we choose the final $\gamma$ values 1, 1, and 0.5, respectively.

\begin{figure}[ht]
	\centering
	\includegraphics[width=.45\textwidth]{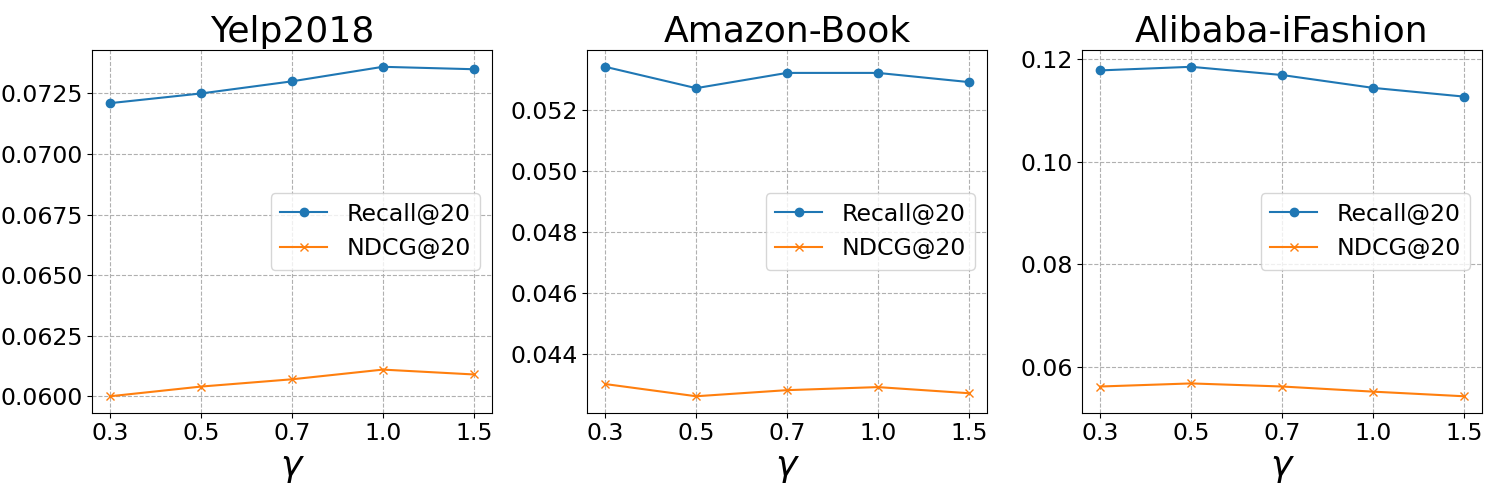}
	\caption{Influence of the magnitude $\gamma$ of TwinCL. }
	\label{fig:gamma}
\end{figure}

\section{Related Work}

\noindent \textbf{GNN-Based Recommender Systems} 
Graph Neural Networks (GNNs)~\cite{wu2020comprehensive} have been extensively employed in collaborating filtering, compared to RNN-based~\cite{hidasi2015session, liu2024behavior} and Transformer-based~\cite{kang2018self, yuan2022multi, sun2019bert4rec} recommendation models. Leveraging graph convolutions~\cite{kipf2016semi, zhang2019graph}, GNNs can aggregate the embeddings of neighboring nodes to enhance the embeddings of the target user or item nodes. This ability makes GNNs particularly adept at link prediction, a foundational task in recommendation systems. GCMC~\cite{berg2017graph} is one of the early studies, which employs graph convolutions on bipartite interaction graphs specifically for link prediction in recommendation contexts. With the popularity of GNNs, an increasing number of collaborative filtering methods have incorporated GNNs~\cite{ying2018graph, sun2019multi, wang2019knowledge, wang2019neural, he2020lightgcn, wang2020disentangled}. For instance, NGCF~\cite{wang2019neural} adapts the Graph Convolutional Network (GCN) directly for collaborative filtering, wherein graph convolutions model high-order connectivity. LightGCN~\cite{he2020lightgcn} streamlines the GCN by eliminating redundant operations, such as transformation matrices and nonlinear activation functions, thereby rendering the GCN more effective for collaborative filtering. Today, many GNN-based recommenders leverage LightGCN for encoding, with this encoding approach serving as a backbone for several graph contrastive learning-based recommendation models. Notable examples of such models that use LightGCN encoders include SGL~\cite{wujc2021self}, NCL~\cite{lin2022improving}, and SimGCL~\cite{yu2022graph}, each applying distinct contrastive learning and augmentation strategies. 

\vspace{1ex} \noindent \textbf{Contrastive Learning in Recommendation}
Contrastive learning~\cite{jaiswal2020survey, khosla2020supervised} has been widely applied in computer vision~\cite{bachman2019learning, chen2020simple}, natural language processing~\cite{gao2021simcse}, and recommender systems~\cite{yao2021self, yu2023self}. Due to the slow convergence and the lack of explicit addressing of the data sparsity issue in NGCF and LightGCN, Graph Contrastive Learning (GCL)~\cite{liu2022graph, wu2021self} has also been introduced into graph collaborative filtering. Graph contrastive learning has emerged as an effective strategy to refine the representations learned by GNNs, often leading to noticeable performance gains. GraphCL~\cite{you2020graph} investigates a variety of graph augmentations, encompassing node dropping, edge perturbation, attribute masking, and subgraph sampling. Building upon this exploration of graph augmentations, SGL~\cite{wujc2021self} studies graph augmentations tailored for collaborative filtering, employing techniques like node dropout, edge dropout, and random walks to derive contrastive views. Furthermore, SimGCL~\cite{yu2022graph} questions the indispensability of graph data augmentations for contrastive learning. Instead, it suggests a simplified graph augmentation technique that involves introducing random noise directly to the feature representation via noise perturbation. Apart from constructing the contrastive pairs by random sampling, NCL~\cite{lin2022improving} introduces a semantic prototype in the semantic space, facilitating the learning of both structural neighbors within graphs and their semantic counterparts. This approach provides inspiration for the design of GCL methods, with a notable emphasis on harnessing both structural and semantic information to enhance recommendation performance. In addition, there are GCL models that integrate knowledge graphs~\cite{ji2021survey} such as KGCL~\cite{yang2022knowledge}, along with models employing dual encoders for data augmentation, like MERIT~\cite{jin2021multi} and SimGRACE~\cite{xia2022simgrace}. These graph models have shown promising results in general graph learning but may not be as efficient for collaborative filtering.

\section{Conclusion}

In this paper, we revisit the domain of graph contrastive learning for recommendation, emphasizing the significance of alignment and uniformity properties. Our novel approach, the Twin Graph Contrastive Learning model, offers a streamlined contrastive learning framework by introducing a straightforward twin encoder while eliminating the need for traditional augmentation techniques in graph contrastive learning. Through its updating mechanism, our model generates diverse contrastive views at different training stages, thereby enhancing the efficacy of contrastive learning. Furthermore, the twin encoder capitalizes on the advantages of optimizing alignment and uniformity, resulting in accelerated convergence and a more stable training process. The comprehensive experimental study demonstrates the advantages of our method in both recommendation performance and training efficiency.

\bibliographystyle{ACM-Reference-Format}
\balance
\bibliography{ref}

\appendix

\end{document}